# Investigating mixed traffic dynamics of pedestrians and non-motorized vehicles at urban intersections: Observation experiments and modelling

Chaojia Yu[1], Kaixin Wang[1], Junle Li[1], Jingjie Wang[1,2*]

[1]Pittsburgh Institute, Sichuan University, Chengdu, 610207, China
[2]School of Industrial Engineering, Purdue University, West Lafayette, IN 47907, United States

*Corresponding author: Jingjie Wang, wang0601@purdue.edu

**Abstract**

Urban intersections with mixed pedestrian and non-motorized vehicle traffic present complex safety challenges, yet traditional models fail to account for dynamic interactions arising from speed heterogeneity and collision anticipation. This study introduces the Time and Angle Based Social Force Model (TASFM), an enhanced framework extending the classical Social Force Model by integrating Time-to-Collision (TTC) metrics and velocity-angle-dependent tangential forces to simulate collision avoidance behaviors more realistically. Using aerial trajectory data from a high-density intersection in Shenzhen, China, we validated TASFM against real-world scenarios, achieving a Mean Trajectory Error (MTE) of 0.154 m (0.77% of the experimental area width). Key findings reveal distinct behavioral patterns: pedestrians self-organize into lanes along designated routes (e.g., zebra crossings), while non-motorized vehicles exhibit flexible path deviations that heighten collision risks. Simulations of three conflict types (overtaking, frontal/lateral crossing) demonstrate TASFM's capacity to replicate adaptive strategies like bidirectional path adjustments and speed modulation. The model provides actionable insights for urban planners, including conflict hotspot prediction and infrastructure redesign (e.g., segregated lanes), while offering a scalable framework for future research integrating motorized traffic and environmental variables. This work advances the understanding of mixed traffic dynamics and bridges the gap between theoretical modeling and data-driven urban safety solutions.

**Keywords:** Mixed traffic flow; Social force model; Pedestrian dynamics; Time-to-collision; Traffic safety.

# 1. Introduction

In urban environments, the interaction between pedestrians and non-motorized vehicles is a significant aspect of traffic dynamics, particularly at intersections where the potential for chaos increases alongside urbanization. These interactions are shaped by distinct behavioral patterns: pedestrians exhibit self-organizing phenomena such as lane formation driven by following and avoidance behaviors (Tajima et al., 2002; Guo et al., 2015), while non-motorized vehicles display heterogeneous maneuvers like overtaking and lateral deviations from designated lanes (Mohammed et al., 2019; Zhang & Ge, 2021). Pedestrian speed and flow are further influenced by demographic factors, with young males moving faster than females or elderly individuals, and luggage-carrying pedestrians exhibiting reduced speeds (Ye et al., 2012). Conversely, non-motorized vehicles maintain an average speed of around 3.5 m/s even under high-density conditions, prioritizing rapid traversal over collision avoidance (Yan et al., 2015; Li et al., 2015). Recognizing the dual objectives of ensuring efficient vehicle flow and safeguarding road user safety, there is a pressing need to delineate these interactions accurately and develop effective management strategies (Fu et al., 2025).

Historically, both theoretical and experimental approaches have been employed to analyze pedestrian dynamics (Cao et al., 2019). Theoretical frameworks like self-organizing behavior and models such as the social force model (Helbing & Molnár, 1995) and heuristics-based models (Moussaïd et al., 2011) have been utilized to describe and quantify traffic phenomena. For instance, Helbing et al. (2000) extended the social force model to simulate panic-driven crowd behavior, while Moussaïd et al. (2011) introduced cognitive heuristics to explain pedestrian decision-making in complex scenarios. Moreover, advancements in technology have enabled researchers to simulate these interactions based on real-world scenarios, with data often gathered via drones (Wang et al., 2022). Traditional experimental methods have leveraged machine learning algorithms like Scale Invariant Feature Transform (SIFT) and Directional Gradient Histogram (HOG) to extract features, subsequently classified by algorithms such as Support Vector Machine (SVM) and AdaBoost (Kazemi et al., 2007; Ma & Grimson, 2005; Taigman et al., 2014). In recent developments, deep learning frameworks such as YOLO v8 and Deepsort have been adapted to train models for precise trajectory detection of pedestrians and non-motorized vehicles (Zhou & Wang, 2024; Du et al., 2022). For example, Narayanan et al. (2021) combined HOG with YOLO for thermal image pedestrian detection.

Despite the prowess of the Social Force Model (SFM) in modeling pedestrian flow since its inception in 1995 (Helbing & Molnár, 1995; Helbing et al., 2000), most studies have predominantly focused on pedestrian-only scenarios. For example, Jiang et al. (2017) extended the SFM to bidirectional pedestrian flows but overlooked interactions with non-motorized vehicles. However, in settings such as intersections in China, non-motor vehicles, characterized by their higher speeds and less predictable maneuvers, form a much more substantial portion of the traffic (Wang et al., 2023). This necessitates considering additional factors like velocity and orientation in force calculations, which traditional models based solely on distance fail to address adequately (Guo et al., 2015; Sheykhfard et al., 2021).

Therefore, it is in need for more comprehensive models that can mirror real-world traffic conditions with both pedestrian and non-motorized vehicle mixed flow (Zhang & Ge, 2021; Xie et al., 2022). To systematically evaluate how existing studies capture the dynamic characteristics of mixed flows, **Table 1** compares key literature across three critical dimensions: interaction mechanisms, behavioral heterogeneity, and model validation methodologies.

Table 1. Relevant literature in mixed traffic dynamics analysis

| Reference | Research Focus | Research Methods | Observation Experiment | Simulation Modelling |
|---|---|---|---|---|
| Wang et al., 2019 | Multi-class detection | CNN object detection | √ | |
| Narayanan et al., 2021 | Thermal pedestrian detection | Hybrid ML & YOLO | √ | |
| Hung et al., 2018 | Obstacle detection | YOLO, hardware deployment | | √ |
| Zhou & Wang, 2023 | Drone traffic tracking | YOLO, trajectory tracking | √ | |
| Tao et al., 2017 | Accelerated traffic detection | YOLO and R-FCN | | √ |
| Wang et al., 2023 | Pedestrian-bicycle flow | Controlled observation | √ | |
| Zhang et al., 2021 | Mixed traffic conflict | Four-stage crossing simulation | | √ |
| Redmon et al., 2018 | Real-time object detection | CNN object detection | √ | |
| Wang et al., 2024 | Real-time object detection | Improved YOLO | √ | |
| **This Paper** | Mixed traffic dynamics and extended social force model | YOLO, Particle swarm optimization, and simulation | √ | √ |

Note: CNN stands for Convolutional Neural Network; ML stands for Machine Learning; YOLO stands for You Only Look Once; R-FCN stands for Region-based Fully Convolutional Network

Current literature reveals several conclusions: (1) most research relies solely on observation experiments or simulation modeling, with limited integration between empirical data and theoretical frameworks; (2) Existing models often treat pedestrians and non-motorized vehicles as homogeneous agents, ignoring non-motorized vehicles' specific behaviors (e.g., lateral maneuverability, speed heterogeneity); (3) while YOLO-based detection has become a prevalent tool for trajectory extraction, its application often lacks scenario-specific adaptations to mixed traffic dynamics. To address these gaps, this study integrates YOLO v8-based observation experiments with a novel simulation framework: a Time and Angle based Social Force Model (TASFM). The TASFM extends the traditional Social Force Model (SFM) by innovatively incorporating Time-to-Collison (TTC) metrics and velocity angle adaptations, enabling dynamic force adjustments that reflect real-world collision risks. Validated against aerial video data from Shenzhen, China, the model not only captures self-organizing phenomena such as lane formation but also bridges empirical observations with theoretical predictions, providing actionable insights into conflict-avoidance mechanisms for heterogeneous traffic participants.

The remainder of this paper is structured as follows: **Section 2** introduces the experimental methodologies and modeling approaches employed; **Section 3** discusses the results and findings; and **Section 4** concludes the paper with future work.

## 2. Methodology

In the study of pedestrian and non-motorized vehicle mixed traffic characteristics at road intersections, this paper proposes a comprehensive methodological framework aimed at accurately extracting and optimizing the trajectories of pedestrians and non-motorized vehicles from real-world video data in Shenzhen, China, and further analyzing the characteristics of mixed traffic. The methodology section of this study encompasses (1) accurately extracting the trajectories of pedestrians and non-motorized vehicles from real-world video, along with analyzing the characteristics of mixed traffic; (2) modelling the traffic dynamics and develop the simulation.

### 2.1 Observation experiment
### 2.1.1 Data collection and preprocessing

The primary source of data for this study is aerial video footage captured by drones (DJI Air 2S) at a busy intersection in Shenzhen, China. The selected videos cover various traffic conditions at different times, ensuring a comprehensive and representative analysis. Preprocessing steps include stabilization, noise reduction, and illumination adjustment to enhance the accuracy and efficiency of subsequent processing. Detailed information for the field experiment is shown in **Fig. 1** and **2**.

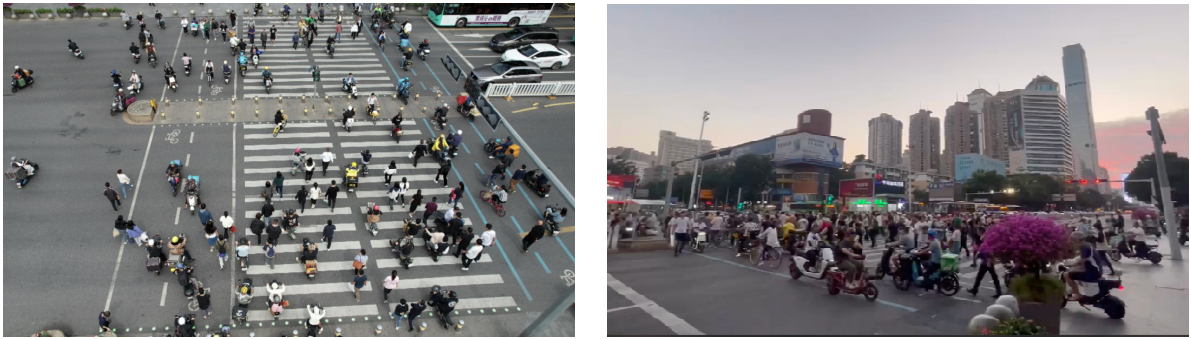

**Fig. 1.** Real road condition of the intersection

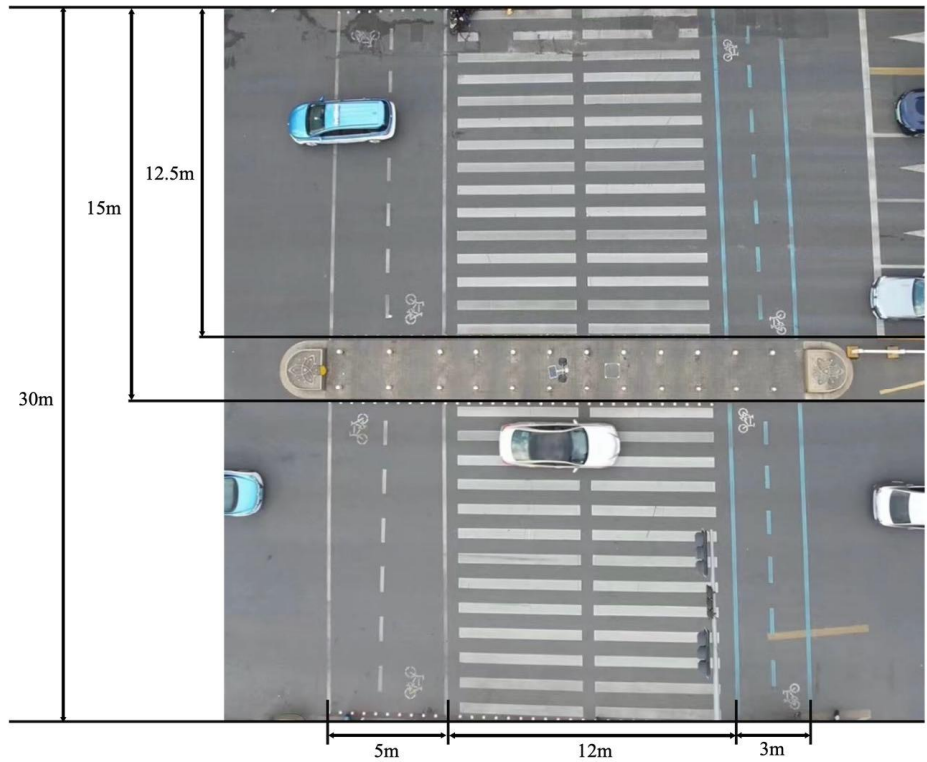

**Fig. 2.** Road dimension of the intersection

### 2.1.2 YOLO algorithm

This study uses YOLO v8 for trajectory extraction. The network of YOLO follows a typical YOLO-style architecture, as presented in **Fig. 3**. It consists of a feature extraction backbone (with C2f and SPPF), a multi-scale fusion neck with up sampling and concatenation, and multiple prediction heads to detect objects at different scales. The methods operate by directly predicting bounding boxes and classes within a single network, hence delivering the final detection results in a rapid and end-to-end manner. Numerous research studies have been conducted by researchers focusing on the application of YOLO in detecting traffic obstacles. YOLO v8 is specifically chosen for its ability to handle real-time video processing, making it ideal for analyzing dynamic traffic scenes with multiple moving objects.

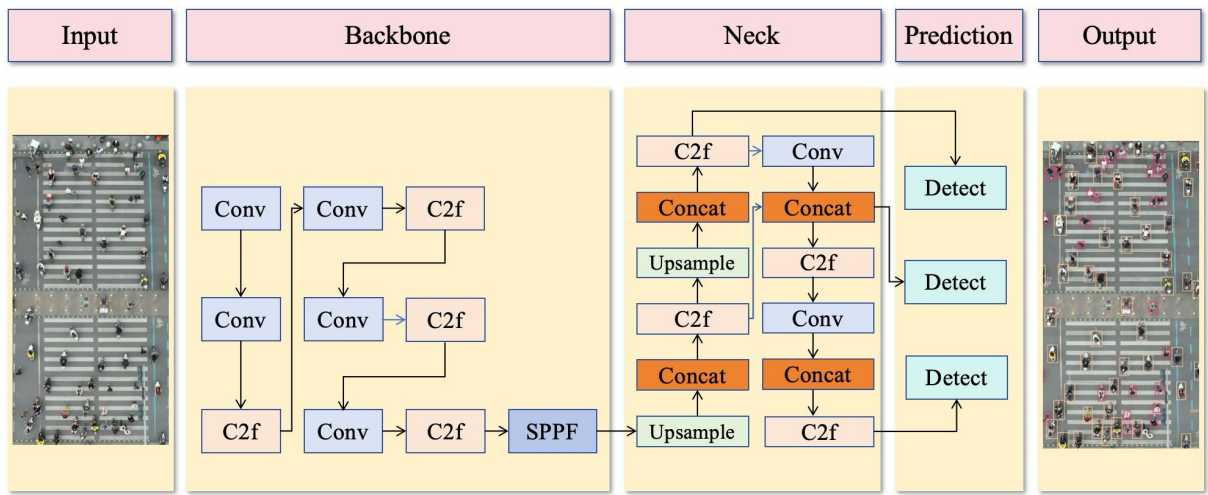

Fig 3. The YOLO algorithm framework diagram

### 2.1.3 Model Training

To ensure the precision of trajectory extraction from the aerial footage, the process involved both manual annotation and the use of YOLO. Initially, the work started from manually annotating the videos, identifying and marking the paths and classification of pedestrians and non-motorized vehicles.

After the manual annotations were completed, this dataset was used to train a detection model using YOLO v8, a highly efficient deep learning framework known for its speed and accuracy in object detection tasks. The model was trained to recognize and track pedestrians and non-motorized vehicles automatically, then applied to this study.

## 2.2 Model description
### 2.2.1 Social force model

The Social Force Model (SFM) was introduced by Helbing and Molnár in 1995 as a novel approach to model pedestrian behavior using analogies from physical forces (Helbing and Molnár, 1995). This model utilizes the analogy of Newtonian mechanics to describe how pedestrians move, comprising three primary forces: a self-motivation force $f_i^0(t)$, which propels an individual toward their target; interaction forces $f_{ij}(t)$ among pedestrians; and interaction forces $f_{iw}(t)$ between pedestrians and fixed obstacles like walls. In all, the SFM is articulated through equations as shown in Eq. (1), setting a more standard for subsequent research into pedestrian dynamics in crowded environments (Helbing et al., 2000).

$$m_i \frac{dv_i(t)}{dt} = f_i^0(t) + \sum_{Evacuees} f_{ij}(t) + \sum_{Obstacles} f_{iw}(t) \qquad (1)$$

Here, the self-motivation force $f_i^0(t)$ is formulated by the following.

$$f_i^0(t) = m_i \frac{v_i^0(t)e_i^0(t) - v_i(t)}{\tau_i} \qquad (2)$$

where $m_i$ is the mass of each pedestrian $i$ $(i = 1, 2, 3, ...)$. $v_i^0(t)$ denotes the quantity of desired speed at time $t$. $e_i^0(t)$ means the unit vector of desired direction at time $t$. $v_i(t)$ is the actual velocity of each pedestrian. And $\tau_i$ is the relaxation time.

The interaction force between pedestrian $i$ and $j$ $(j = 1,2,3, ..., j \neq i)$ is defined as Eq. (3) below.

$$f_{ij}(t) = \left\{ A_i exp\left[\frac{(r_{ij} - d_{ij})}{B_i}\right] + Kg(r_{ij} - d_{ij}) \right\} n_{ij} + kg(r_{ij} - d_{ij})\Delta v_{ij}^t t_{ij} \qquad (3)$$

where $A_i$ represents the interaction strength, while $B_i$ indicates the range of interaction. The term $r_{ij}$ is defined as the sum of the radii $r_i$ and $r_j$, and $d_{ij}$ represents the distance between the centers of mass of two pedestrians. The constants $K$ and $k$ are used to denote large numerical values. The vector $n_{ij}$ is the normalized vector that points from pedestrian $j$ to pedestrian $i$, and $t_{ij}$ is the vector representing the tangential direction. The term $\Delta v_{ij}^t$ describes the difference in tangential velocity between pedestrians $j$ and $i$. The function $g(x)$ is defined to be zero when the pedestrians are not in contact (i.e., $d_{ij} > r_{ij}$), and it takes the value of its argument $x$ otherwise.

Moreover, the interaction between a pedestrian $i$ and the obstacle $w$ can be written as Eq. (4).

$$f_{iw}(t) = \left\{ A_i exp\left[\frac{(r_i - d_{imw})}{B_i}\right] + Kg(r_{ij} - d_{imw})\right\} n_{iw} - kg(r_i - d_{imw})(v_i(t) \cdot t_{iw})t_{iw} \quad (4)$$

where $A_i$ and $B_i$ are still treated as constants, with $r_i$ indicating the radius of pedestrian $i$. The variable $d_{imw}$ measures the direct distance between a pedestrian's center of mass and any wall or obstacle. The constants $K$ and $k$ still denote large numerical values. The vector $n_{iw}$ is the normalized vector that points from an obstacle or wall towards pedestrian $i$, and $t_{iw}$ represents the tangential direction relative to this. The term $v_i(t)$ specifies the actual velocity of the pedestrian at time $t$. The function $g(x)$ is defined such that it equals zero if there is no contact between the pedestrian and the wall or obstacle (i.e., $d_{iw} > r_i$) and takes the value of $x$ when contact occurs.

By providing a structured and quantifiable approach to understanding pedestrian dynamics, the SFM serves as an indispensable foundation for modeling interactions of people. However, the SFM primarily considers the effects of spatial factors, may overlooking the dynamic aspects of pedestrian motion such as velocity and direction changes. This limitation hinders its applicability to scenarios where interactions are highly dynamic and influenced by factors beyond mere proximity, particularly in mixed traffic environments involving faster-moving non-motorized vehicles. Therefore, adapting the SFM to incorporate additional variables such as velocity and direction can enhance its applicability and precision in urban planning and safety assessments. This paper leverages the SFM's robust framework to develop an enhanced model that addresses the unique challenges posed by the interaction of pedestrians and non-motorized vehicles, aiming to provide a more comprehensive understanding of urban traffic dynamics.

**2.2.2 Time and angle based social force model**

Firstly, a significant enhancement in the extended model is the introduction of a tangential repulsive force component alongside the conventional normal (radial) repulsive force. Unlike traditional models where repulsion is only considered in the direct line of interaction, the TASFM includes a tangential force that acts perpendicular to this line. This force is not merely an additional component but is dynamically linked with both the normal and tangential repulsive forces, with its magnitude dependent on the angle $\theta$ of velocity between two interacting bodies as shown in **Fig. 4** and Time-to-Collision (TTC) in **Fig. 5**.

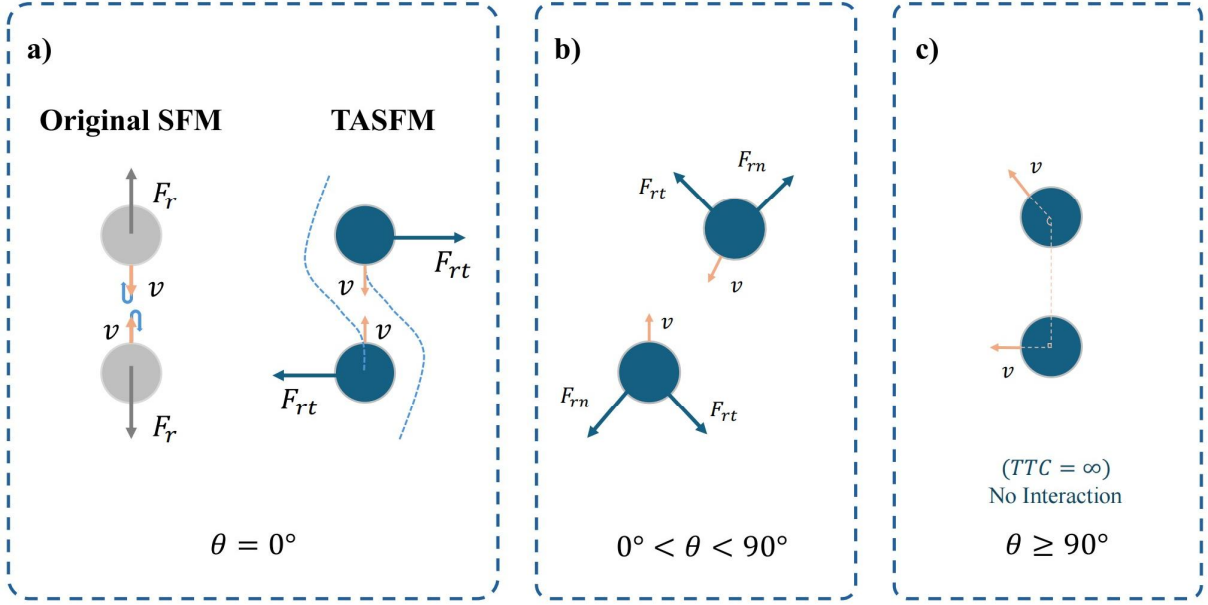

**Fig. 3.** Interaction dynamics in Original SFM and TASFM with the influence of angle

To achieve this, two multipliers (i.e., $sin\theta$ and $-cos\theta$) are then added into the interaction force $f_{ij}(t)$ in Eq. (3) as follows.

$$f_{ij}(t) = A_i exp(B_i \Delta t)[n_{ij} sin\theta + t_{ij}(-cos\theta)] + Kg(\Delta t)n_{ij} + kg(\Delta t)\Delta v_{ij}^t t_{ij} \qquad (5)$$

In practical terms, this means that if two entities are heading directly towards each other (head-on collision), both the tangential and normal force will encourage lateral movement to avoid collision based on the angles, instead running into each other. The interaction between these tangential and normal forces allows for a more nuanced and effective dispersal of entities, facilitating smoother flow and avoiding potential collisions by allowing entities to sidestep each other rather than stopping or reversing direction.

Moreover, Time-to-Collision (TTC) is calculated based on the relative speed and trajectory of the moving entities, allowing the model to dynamically adjust the social forces according to how imminent a collision is perceived to be. To determine the possibility and timing of a collision between two entities, such as particles $i$ and $j$, one must consider their positions and velocities at a given time $t$. If we denote the positions of particles $i$ and $j$ at the moment of potential contact (i.e., $t + \Delta t$), by $(r'_{xi}, r'_{yi})$ and $(r'_{xj}, r'_{yj})$ respectively, the critical separation distance at which they collide is given by $\sigma = \sigma_i + \sigma_j$. Mathematically, this condition is expressed as:

$$\sigma^2 = (r'_{xi} - r'_{xj})^2 + (r'_{yi} - r'_{yj})^2 \qquad (6)$$

Considering that the particles move along straight-line trajectories leading up to the collision, the future positions can be projected as:

$$r'_{xi} = r_{xi} + \Delta t \cdot v_{xi} \quad (7)$$

$$r'_{xj} = r_{xj} + \Delta t \cdot v_{xj} \quad (8)$$

$$r'_{yi} = r_{yi} + \Delta t \cdot v_{yi} \quad (9)$$

$$r'_{yj} = r_{yj} + \Delta t \cdot v_{yj} \quad (10)$$

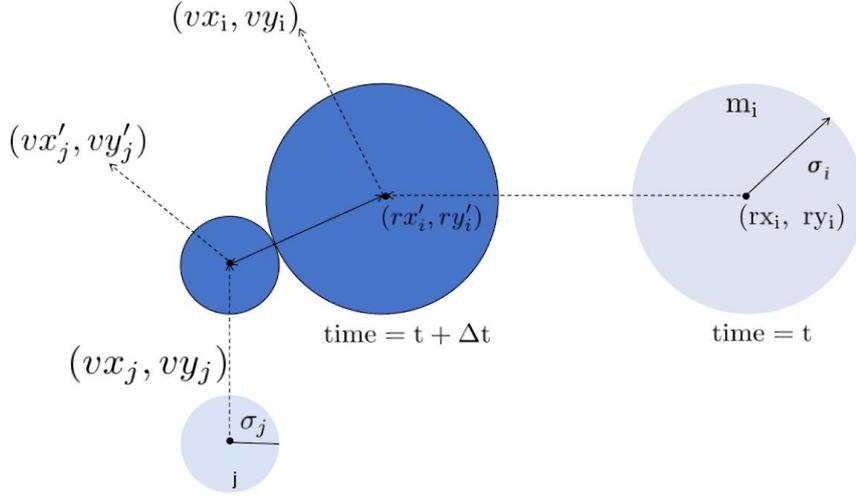

**Fig. 5.** Illustration for the Time-to-Collision between two objects

By substituting these expressions back into the equation for $\sigma^2$ and solving the resulting quadratic equation for $\Delta t$, the physically relevant root is selected. Simplifying this solution provides an explicit formula for $\Delta t$ in terms of the known positions, velocities, and radii of the particles as follows.

$$\Delta t = \begin{cases} \infty, & \text{if } \Delta v \cdot \Delta r \geq 0 \\ \infty, & \text{if } d < 0 \\ -\dfrac{\Delta v \cdot \Delta r + \sqrt{d}}{\Delta v \cdot \Delta r}, & otherwise \end{cases} \quad (11)$$

where $d = (\Delta v \cdot \Delta r)^2 - (\Delta v \cdot \Delta r)(\Delta r \cdot \Delta r - \sigma^2)$

Hence, it allows the TASFM to accurately predict collision times and adjust the interactive forces $f_{ij}(t)$ in Eq. (3), accordingly.

$$\boldsymbol{f}_{ij}(t) = A_i \exp\left[\frac{-\Delta t}{B_i}\right] \boldsymbol{n}_{ij} \sin\theta + C_i \exp\left[\frac{-\Delta t}{D_i}\right] \boldsymbol{t}_{ij}(-\cos\theta) \\ + k \cdot g(r_{ij} - d_{ij})\boldsymbol{n}_{ij} + \kappa \cdot g(r_{ij} - d_{ij})\Delta v^t_{ij}\boldsymbol{t}_{ij} \quad (12)$$

The original SFM has traditionally modeled pedestrian interactions primarily through spatial considerations, notably the distance between entities. This method, while foundational, occasionally results in unrealistic simulations of avoidance behaviors, where repulsive forces may be exerted even when two objects are not on a collision course. Such scenarios can lead to exaggerated or unnecessary deviations in the paths of pedestrians or vehicles, detracting from the realism and applicability of the model in practical urban settings. Therefore, to address these limitations, our enhanced model as shown in **Fig. 6** introduces a conditional aspect: if the calculated TTC indicates that no collision is imminent (i.e., the entities will not collide if they maintain their current paths), no repulsive force is applied. Conversely, the closer two objects are to a potential collision, the greater the repulsive force exerted between them.

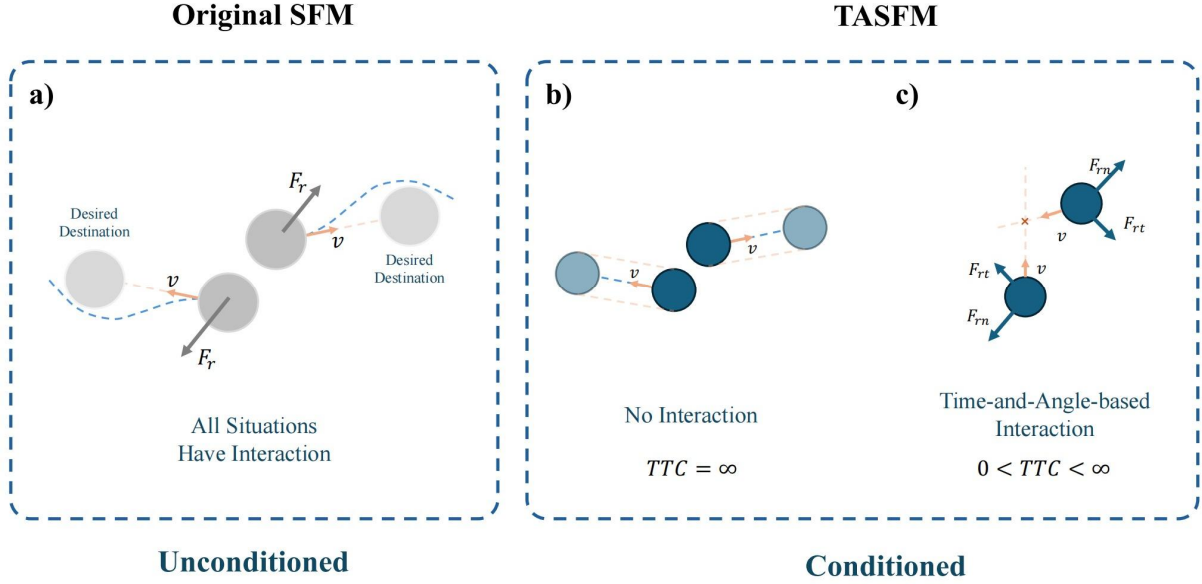

**Fig. 6.** Interaction dynamics in original SFM and TASFM with the influence of TTC

This modification ensures that the model's responses are more grounded in the actual dynamics of the situation, providing a more realistic simulation of pedestrian and non-motorized vehicle interactions. The forces now not only consider the immediate spatial configuration but also the projected future states based on velocity and trajectory. This leads to a more precise and contextually appropriate application of forces, effectively reducing unnecessary avoidance maneuvers and enhancing the model's utility for planning and analysis in mixed traffic environments.

Building on these modifications, TASFM provides a sophisticated framework for analyzing and predicting the movement patterns of pedestrians and non-motorized vehicles at intersections. By considering both the time to potential collisions and the angles of interaction, TASFM offers a more detailed and realistic simulation of social forces in high-density urban environment and helps in understanding how individuals intuitively negotiate space and avoid collisions in real-world scenarios.

Lastly, this paper uses PSO (Particle Swarm Optimization) algorithm to precisely determine the parameters in Eq. (11) based on the real data extracted in the observation experiment. The following pseudocode is the illustration for the algorithm, where $\theta$ denotes the set of parameters. To measure the error in simulation, Mean Trajectory Error (MTE) has also been applied. The metric is elaborated in **Section 3.2.2**.

**Algorithm: Particle Swarm Optimization (PSO)**

1. Initialize N particles with $\theta_i, v_i$ randomly
2. Evaluate the fitness of $\theta_i$, set $p\_best_i \leftarrow \theta_i$
3. Set global best $g\_best \leftarrow$ best among $p\_best_i$

For $t = 1$ to max_iterations:

For each particle $i$:
$$v_i \leftarrow w \cdot v_i + c_1 \cdot r_1 \cdot (p\_best_i - \theta_i) + c_2 \cdot r_2 \cdot (g\_best_i - \theta_i)$$
$$\theta_i \leftarrow \theta_i + v_i$$
If the fitness of $\theta_i$ better than the fitness of $p\_best_i$:
$$p\_best_i \leftarrow \theta_i$$
$g\_best \leftarrow$ best among $p\_best_i$

Return $g\_best$

## 3. Results and analysis
### 3.1 Analysis of observation experiment
#### 3.1.1 Trajectory extraction

The aerial footage captured at a bustling intersection in Shenzhen is meticulously analyzed using the YOLO deep learning model for object detection and trajectory extraction. The evaluation of YOLO's performance during the validation phase yielded promising results as shown in **Fig. 7** below, underscoring its effectiveness in traffic analysis applications. In a validation set consisting of multiple images, the YOLO model detected a total of 466 instances, split between 242 non-motorized vehicles and 224 pedestrians. The model achieved a precision of 0.868 and a recall of 0.889, with an mAP50 of 0.881, indicating high accuracy in identifying relevant traffic participants. These metrics are particularly important for ensuring the reliability of the traffic flow data derived from the model's output.

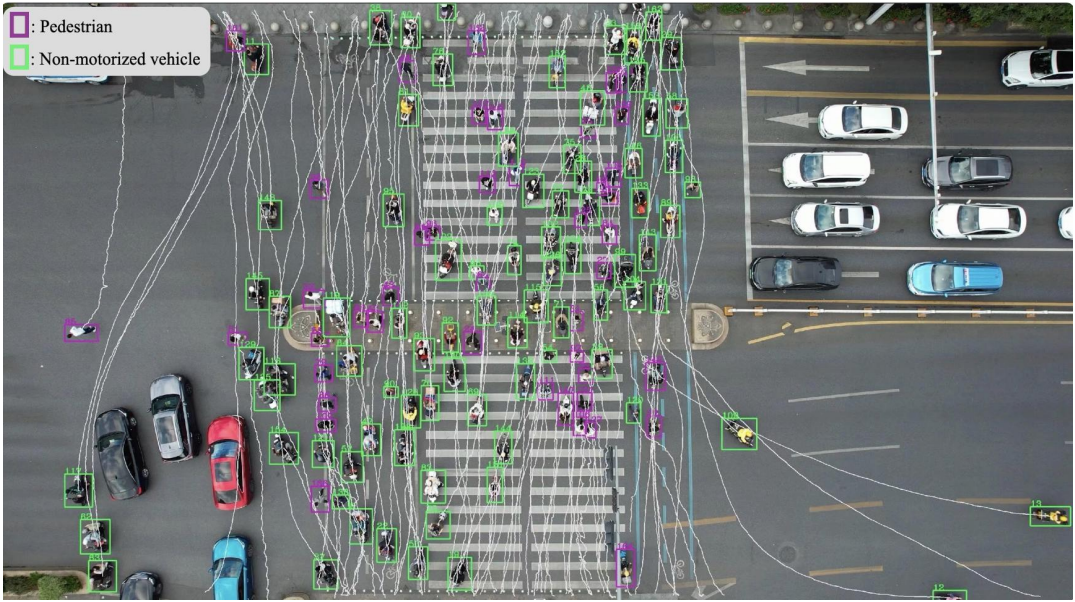

**Fig. 7.** Pedestrian and non-motor vehicle classification and trajectory extraction

#### 3.1.2 Characteristics of the mixed traffic

The study of dynamic parameters in mixed traffic scenarios, particularly concerning non-motorized vehicles and pedestrian interactions, provides critical insights into urban traffic management and planning. This section further discusses the empirical analysis derived from three key visual data representations, examining the impacts of various factors on the speed and movement patterns of non-motorized vehicles within different pedestrian density levels.

The speed distributions of pedestrians and non-motorized vehicles elucidate distinct speed profiles for each group, as shown in **Fig. 8. (a)** and **(b)**. The speed distribution of pedestrians exhibits a sharper peak around 1.2 m/s and declines rapidly beyond 1.7 m/s, indicating that most pedestrians move at relatively low and consistent speeds. In contrast, non-motorized vehicles display a broader distribution with a notable peak around 2.0 m/s and a longer tail extending beyond 4.0 m/s, showing greater variability and generally higher travel speeds. These insights about the characteristics of both groups can be utilized in the model to crucially enhance simulation performance.

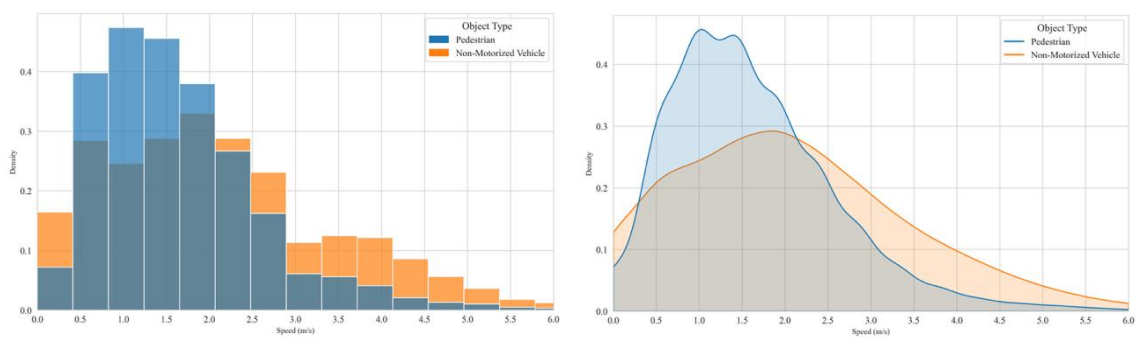

**Fig. 8. (a)** Observed speed distribution histogram and **(b)** kernel density estimation of speed distributions for pedestrians and non-motorized vehicles

The YOLO model has also successfully captured the line formation process, as shown in the heat maps in **Fig. 9. (a)** and **(b)**. In **Fig. 9. (a)** The heat map shows three main routes that pedestrians tend to choose, indicating the formation of lanes. In other words, pedestrians tend to follow the routes that some leaders have walked in front of them, and such self-organized behavior will then lead to the lane formation phenomenon. On the contrary, in **Fig. 9. (b)**, the heat map shows a vaguer preference for non-motorized vehicles when selecting their routes. Therefore, it is evident that pedestrians have less flexibility choosing their routes and tend to choose to walk along the zebra crossing, which is the route designated for pedestrians. However, non-motorized vehicles have a wider range of route selections, and don't usually follow the designated area (the left lane and the right lane next to the zebra crossing area). A great portion of non-motorized vehicles have crossed through the left side of the zebra crossing to the right side or have crossed through conversely. This is a phenomenon that should be meticulously paid attention to because non-motorized vehicles tend to deviate from their designated lanes and easily cause conflict with pedestrians, causing a high risk of clash and even accidents.

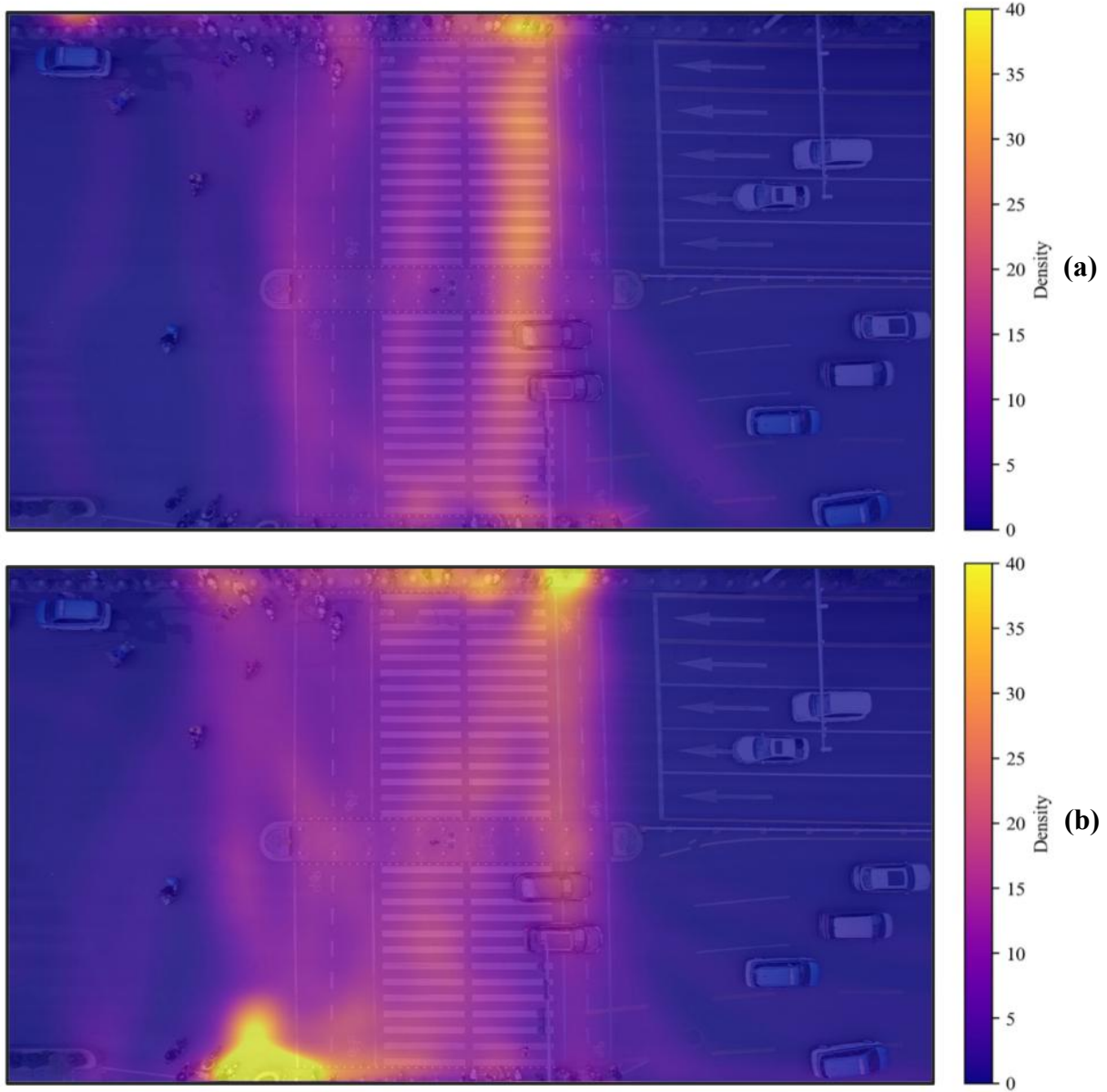

**Fig. 9.** Trajectory heat map for **(a)** pedestrians and **(b)** non-motorized vehicles with density

To better examines the interaction between pedestrian and non-motorized vehicles, the impact of pedestrian density on the speed of non-motorized vehicles is analyzed as shown in **Table 2** and **Fig. 10** below. Pedestrian density refers to global density here (number of people/area, unit: pers/$m^2$), categorized by four different level, i.e., level 1- (0.000, 0.027], level 2- (0.027, 0.040], level 3- (0.040, 0.130], and level 4- (0.130, 0.183]. **Fig. 10** shows a clear trend that the progression from pedestrian level 1 through level 4 reveals a clear trend: as pedestrian density increases, the speed range of non-motorized vehicles tends to shrink. Initially, at lower densities, non-motorized vehicles exhibit a wide range of speeds, likely due to the greater availability of space and fewer interactions requiring speed adjustments. In such low-density situations (level 1 and 2), non-motorized vehicles would likely to reach higher speed like 4 m/s. Low-density situation not only has a high speed upper-bound, but also a low speed lower-bound. This difference is likely caused by the different positions in the crossroad.

Low-density areas are more likely to be not designated to pedestrians. Also, as discussed in the heat map, pedestrians tend to walk only through the zebra crossing. Therefore, low-density areas also include the areas that are not necessary for crossing the road, and the non-motorized vehicles in such areas are also not necessarily crossing the road. They are probably just waiting or resting, causing the lower bound and the wide speed range.

Furthermore, as density increases (level 3 and 4), the operational space reduces, non-motorized vehicles would have less chance to reach a high speed like 4 m/s, causing the upper-bound to abate. However, non-motorized vehicles tend to maintain a mean speed of about 3.5 m/s even in high pedestrian density level situation, indicating their lack of safety awareness. The increase of the lower-bound is likely due to the non-motorized-vehicle-specific demand of passing the high-density area more rapidly and maintain the high speed. Under such a demand, they tend to choose a moderate high speed to cross the high-density area in a shorter time without apparent decreasing their speed.

**Table 2.** Speed of non-motorized vehicles among different pedestrian density levels

| Pedestrian density level | Speed of non-motorized vehicles (m/s) | | | | |
|---|---|---|---|---|---|
| | Minimum | 25 Percentile | 50 Percentile | 75 Percentile | Maximum |
| Level 1 | 0.00 | 2.79 | 3.49 | 4.07 | 6.85 |
| Level 2 | 1.39 | 3.13 | 3.69 | 4.45 | 6.75 |
| Level 3 | 2.16 | 2.90 | 3.19 | 3.53 | 4.50 |
| Level 4 | 2.75 | 3.29 | 3.47 | 3.67 | 4.40 |

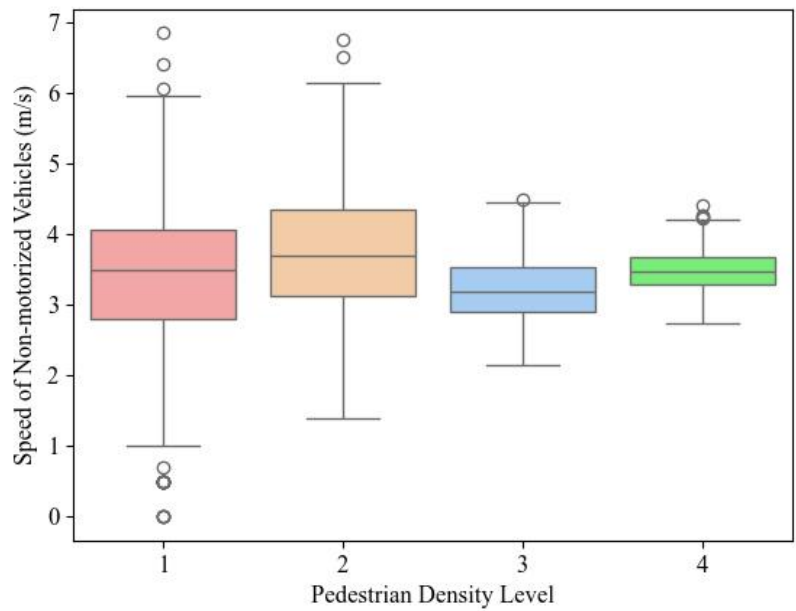

**Fig. 10.** Impact of pedestrian density on the speed of non-motorized vehicles

In short, these figures paint a comprehensive picture of the dynamic interactions between pedestrians and non-motorized vehicles in urban settings. The analyses indicate that while non-motorized vehicles are capable of higher speeds, they tend to maintain a speed of 3m/s and the operational speeds are influenced by pedestrian density, and they usually lack the safety awareness, necessitating adaptive strategies in urban design and traffic regulation.

## 3.2 Model simulation and verification
### 3.2.1 Simulation of typical conflict interaction with TASFM

Traffic conflicts in mixed traffic can be classified based on the angle between pedestrians and non-motorized vehicles at the time of conflict. The first type of interaction, "passing from behind" or "overtaking", involves a trajectory of plus or minus 30 degrees in the same direction. The second type of interaction, known as "crossing from the front", is characterized by trajectories that intersect by positive or negative 30 degrees in opposite directions. Type 3, "crossing from the side", includes all other interactions except the two former types (Beitel et al., 2018). The three specific types of traffic conflicts are shown in **Fig. 11** below.

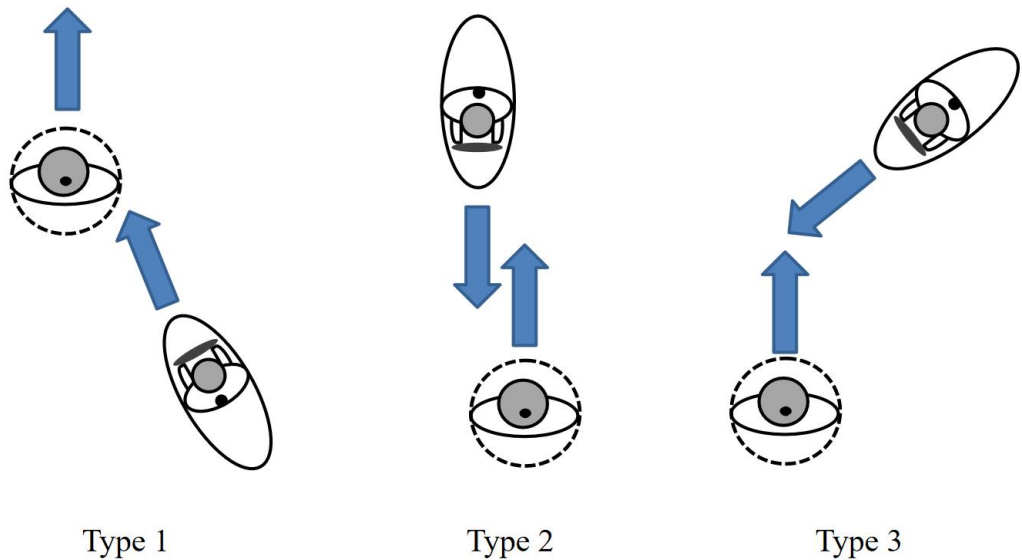

**Fig. 11.** Three types of interaction for the pedestrians and non-motorized vehicles traffic conflicts

To better analyze the conflict avoidance behavior, we first select several typical conflict pairs and utilize TASFM to conduct the simulation. With the best parameters of the model acquired from PSO and by setting the initial position, final position, initial velocity, and object types, we simulate the trajectories of different object pairs under various interaction scenarios. These simulated trajectories are then compared with the real ones extracted from the video, as shown in **Fig. 12**. In the figure, simulated trajectories are shown in darker colors, while real trajectories are shown in lighter colors. Brighter colors indicate higher speeds.

The results in **Fig. 12** below show that TASFM can effectively simulate trajectories in different situations. To gain a deeper understanding of the collision avoidance behavior of pedestrians and non-motorized vehicles, we further construct three typical conflict scenarios and simulate them using TASFM.

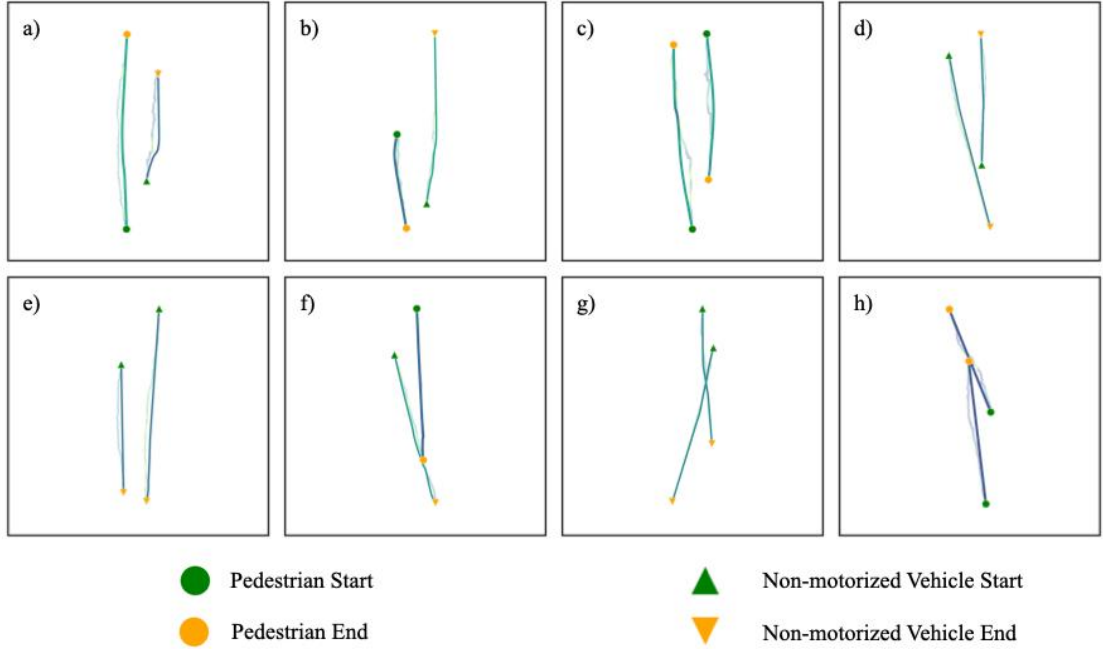

**Fig. 12.** Conflict pairs with simulated and actual trajectories

In the overtaking scenario (i.e., Type 1) as shown in **Fig. 12 (b)**, **(c)**, and **(d)**, the simulation highlights the mutual path adjustment to ensure the safe interaction. The pedestrian slightly turned counterclockwise, and the non-motorized vehicle slightly turned clockwise to prolong the time-to-collision and provide extra time for the non-motorized vehicle to cross the pedestrian. In the crossing-from-the-front scenario (i.e., Type 2) as shown in **Fig. 13 (b)**, **(c)**, and **(d)**, pedestrian and non-motorized vehicle simultaneously self-organized their movement to escape from each other. After the collision risk was dismissed, they adjusted their direction and went straight line toward the destination. In the crossing-from-the-side scenario (i.e., Type 3) as shown in **Fig. 14 (b)**, **(c)**, and **(d)**, the simulation shows the reciprocal avoidance behavior taken to reduce the risk of collision. The pedestrian and the non-motorized vehicle slightly turned in different directions to avoid the final collision. These maneuvers are critical in preventing collisions, particularly in high-density environments where space and reaction time are limited.

Combined with the non-motorized vehicles' speed analysis, the simulations show that despite the safety awareness problem that non-motorized vehicles have, the interactions between them are relatively safe due to the agile avoidance behaviors from both sides. These adjustments can also demonstrate the model's capacity to dynamically predict and simulate the trajectories based on the immediate context and predicted movements of other road users. To further investigate the ability of TASFM within more complex scenarios, we employ it in a more chaotic situation and conduct the error test to verify its capability.

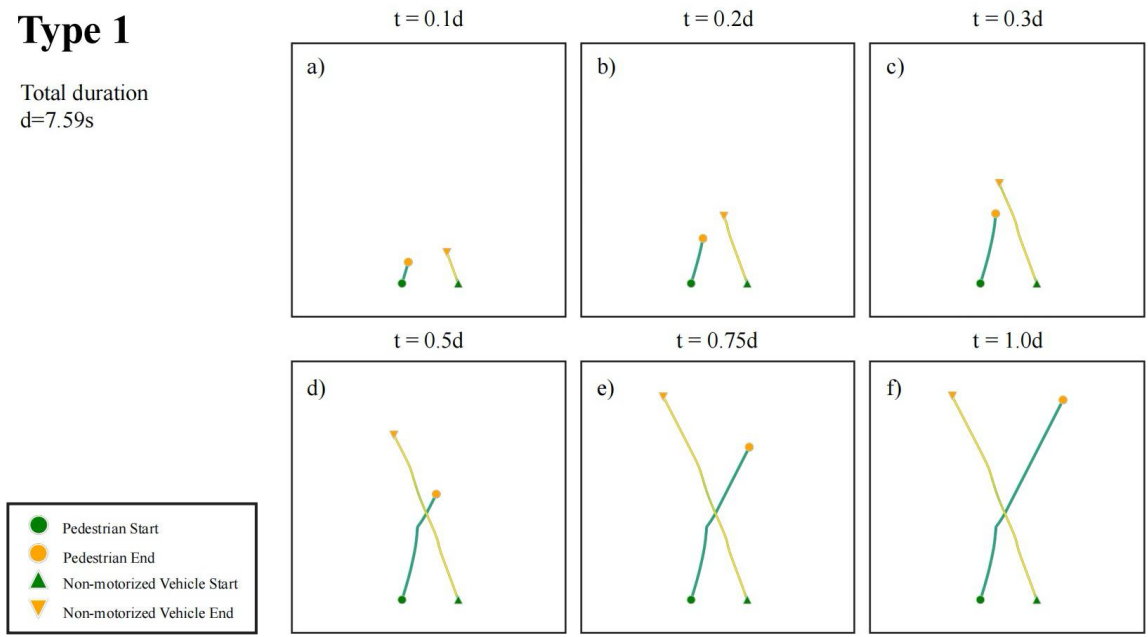

**Fig. 12.** Type 1 conflict pair simulated trajectory

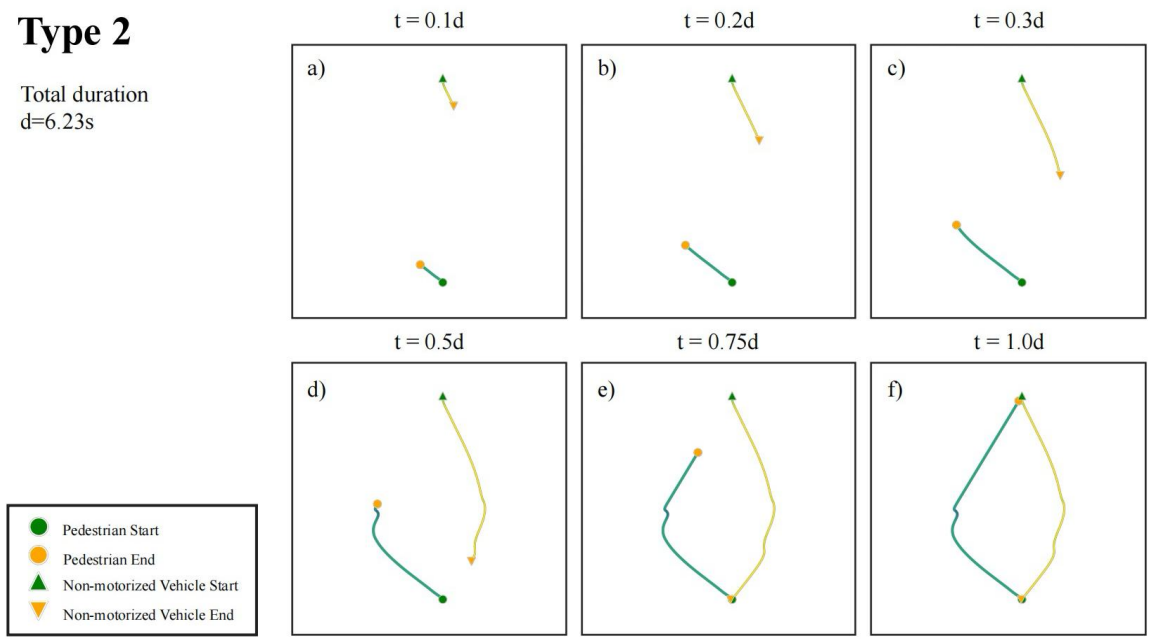

**Fig. 13.** Type 2 conflict pair simulated trajectory

**Type 3**

Total duration
d=6.15s

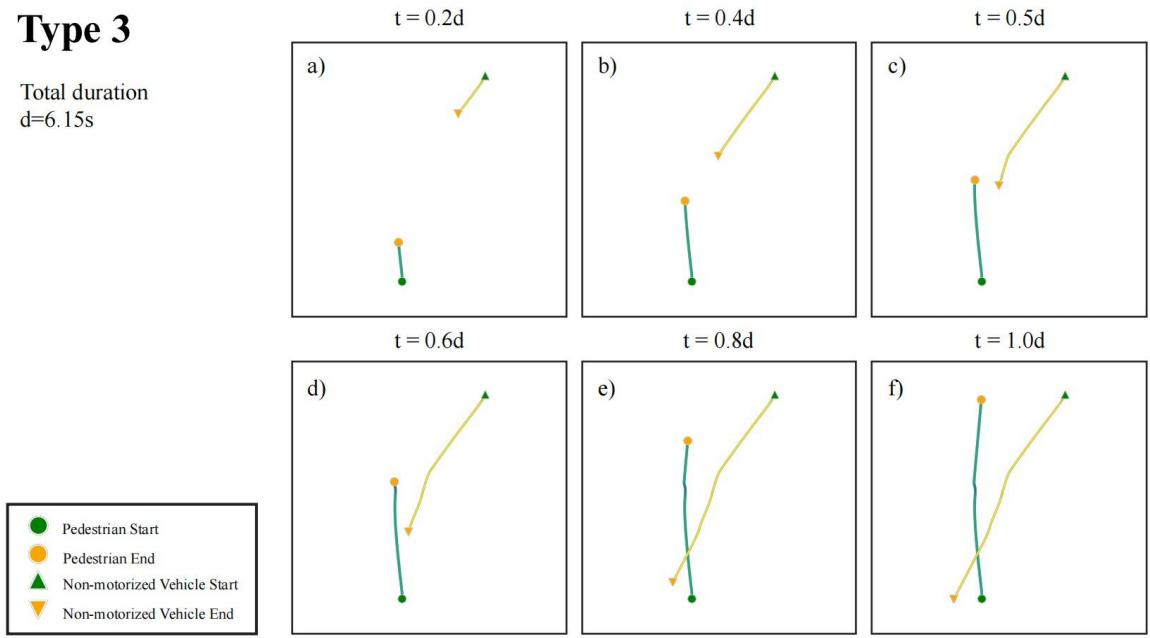

Fig. 14. Type 3 conflict pair simulated trajectory

### 3.2.2 Model verification with error test

By selecting a more complex condition of great disorder, we conduct the TASFM verification, as shown in **Fig. 15**. This visualization is crucial for demonstrating the accuracy and robustness of the simulation model employed. The proximity of the line trajectories to the actual data suggests that the TASFM can effectively replicate real-world dynamics even in more complex situations, capturing the essential trends and deviations with high fidelity. This detailed alignment between the simulated trajectories and the actual data points strongly supports the argument that TASFM is not only reliable but also versatile, adapting accurately to varying conditions, which is indicative of a well-validated and robust simulation framework.

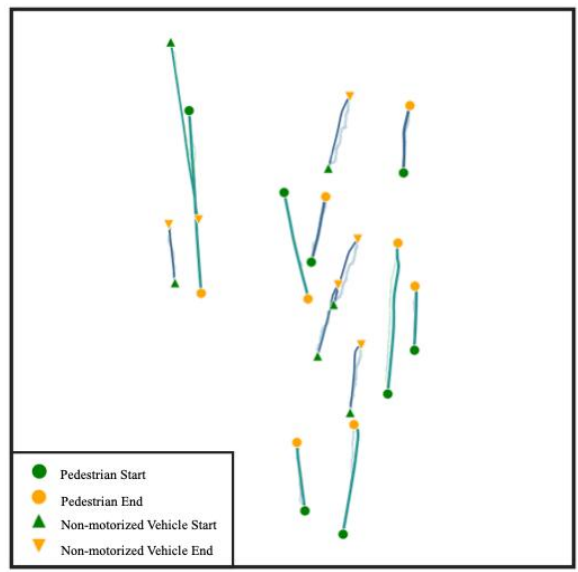

**Fig. 15.** Complex scenario with simulated and actual trajectories

Moreover, to conduct the error test for TASFM, Mean Trajectory Error (MTE) is mainly utilized. The MTE is computed as follows (Zhao et al., 2020).

$$MTE = \frac{1}{total\,TS} \sum_{t=1}^{total\,TS} TE_t = \frac{1}{total\,TS} \sum_{t=1}^{total\,TS} [(x_t, y_t)_{exp} - (x_t, y_t)_{sim}] \quad (13)$$

where $total\,TS$ denotes all timesteps from the simulation, $TE_t$ denotes the positional error at each time step $t$, calculated as the distance between the experimentally observed and the TASFM simulated positions of a pedestrian or non-motorized vehicles, specifically $(x_t, y_t)_{exp} - (x_t, y_t)_{sim}$.

The error for each object per timestep, which is also the mean of the MTE, is calculated as Eq. (14) below.

$$\overline{MTE} = \frac{1}{N} \sum_{i=1}^{N} MTE \quad (14)$$

where $N$ denotes the number of objects.

Through calculation, the mean MTE is 0.154 m, representing only around 0.77% of the total width length (20m) of this field experiment (shown in **Fig. 2**). This result indicates that TASFM not only performs well in simulate dynamic movements but does so with minimal deviation from actual observed trajectories.

Overall, these simulations and error test validate TASFM's utility in enhancing traffic safety by providing realistic anticipations of potential conflicts and enabling proactive adjustments to trajectories. Such capabilities are crucial for developing smarter and safer urban traffic systems where interactions between different types of road users are frequent and varied. These visualizations not only reinforce the robustness of TASFM in simulating real-world scenarios but also offer valuable insights into improving traffic management strategies and infrastructure planning to accommodate the mixed dynamics of pedestrians and non-motorized vehicles.

## 4. Conclusion

This study addressed the critical challenge of modeling mixed pedestrian and non-motorized vehicle interactions at urban intersections, where traditional models often overlook dynamic factors such as velocity variations and collision anticipation. By proposing the Time and Angle Based Social Force Model (TASFM), we extended the classical Social Force Model to incorporate Time-to-Collision (TTC) metrics and tangential repulsive forces influenced by velocity angles. This enhancement enabled a more realistic simulation of conflict-avoidance behaviors, particularly in high-density scenarios where non-motorized vehicles exhibit faster speeds and unpredictable maneuvers.

The practical implications of TASFM are significant for urban planners and policymakers. By predicting conflict hotspots and quantifying self-organizing phenomena like lane formation, the model supports infrastructure redesign (e.g., segregated lanes) and dynamic traffic management systems. However, this study has several limitations. Firstly, the model's validation relied on data from a single intersection in Shenzhen, which may limit its generalizability to other urban contexts with differing traffic norms or infrastructure designs. Second, trajectory extraction using YOLO, while effective and accurate, may still introduce detection inaccuracies in highly congested or occluded scenarios. Additionally, the current framework does not account for environmental factors (e.g., weather, lighting) or behavioral heterogeneity among road users (e.g., age-related speed differences, risk perception).

Future research should integrate multi-city datasets to enhance model robustness and explore interactions with motorized vehicles. Further refinements could incorporate real-time environmental variables and behavioral surveys to capture cultural or demographic influences on decision-making. As cities worldwide grapple with the complexities of multimodal transportation, the models and findings of this paper can help provide a foundational framework for designing safer, more adaptive urban ecosystems that harmonize human behavior, technological innovation, and sustainable mobility goals.

**Acknowledgement**

This research did not receive any specific grant from funding agencies in the public, commercial, or not-for-profit sectors.

**Data availability statement**

The data that support the findings of this study are available from the corresponding author upon reasonable request.

**Competing interest declaration**

The authors declare that they have no known competing financial interests or personal relationships that could have appeared to influence the work reported in this paper.